\documentclass[aps,prx,twocolumn,showpacs,superscriptaddress,floatfix,nofootinbib,longbibliography]{revtex4-1}

\usepackage{graphicx}
\usepackage{subfigure}
\usepackage{natbib,hyperref}
\usepackage{dcolumn}
\usepackage{appendix}

\hypersetup{
  colorlinks   = true, 
  urlcolor     = red, 
  linkcolor    = red, 
  citecolor   = red 
}

\newcommand{\beq}{\begin{equation}}
\newcommand{\eeq}{\end{equation}}
\newcommand{\bea}{\begin{eqnarray}}
\newcommand{\eea}{\end{eqnarray}}

\begin{document}

\title{Topological Crystalline Insulator Nanomembrane with Strain-Tunable Band Gap}

\author{Xiaofeng Qian}
\affiliation{Department of Nuclear Science and Engineering and Department of Materials Science and Engineering, Massachusetts Institute of Technology, Cambridge, Massachusetts 02139, USA}
\author{Liang Fu}
\email[Electronic address: ]{liangfu@mit.edu}
\thanks{corresponding author.}
\affiliation{Department of Physics, Massachusetts Institute of Technology, Cambridge, MA 02139, USA}
\author{Ju Li}
\email[Electronic address: ]{liju@mit.edu}
\thanks{corresponding author.}
\affiliation{Department of Nuclear Science and Engineering and Department of Materials Science and Engineering, Massachusetts Institute of Technology, Cambridge, Massachusetts 02139, USA}
\date{\today}

\begin{abstract}
The ability to fine-tune band gap and band inversion in topological materials is highly desirable for the development of novel functional devices. Here we propose that the electronic properties of a free-standing nanomembrane of topological crystalline insulator (TCI) SnTe and Pb$_{1-x}$Sn$_x$(Se,Te) are highly tunable by engineering elastic strain and controlling membrane thickness, resulting in tunable band gap and giant piezoconductivity. Membrane thickness governs the hybridization of topological electronic states on opposite surfaces, while elastic strain can further modulate the hybridization strength by controlling the penetration length of surface states.  We propose a frequency-resolved infrared photodetector using force-concentration induced inhomogeneous elastic strain in TCI nanomembrane with spatially varying width. The predicted tunable band gap accompanied by strong spin-textured electronic states will open up new avenues for fabricating piezoresistive devices, thermoelectrics, infrared detectors and energy-efficient electronic and optoelectronic devices based on TCI nanomembrane.
\end{abstract}
\pacs{71.70.Ej, 73.20.-r, 07.57.Kp}

\maketitle

\section{Introduction}
\label{introduction}
The ability to tune band gap and control band inversion in topological materials is desirable for the development of novel functional devices. Such band gap engineering has been demonstrated for Bi-based topological insulators using various methods such as alloying \cite{Xu11, Wu13}, thickness engineering \cite{Zhang10, Kim12, Taskin12} and magnetic proximity effect \cite{Xu12b, Checkelsky12, Chang13, Wei13}.  A useful and potentially advantageous alternative approach is elastic strain engineering \cite{Li14}. While there have been extensive theoretical studies on strain tunable band gap in topological insulators \cite{Young11, Bahramy12, Yang12, Kim13, Agapito13, Winterfeld13, Zaheer13, Zhang13},  to our knowledge device concept utilizing this effect has not been demonstrated yet. 

The recently discovered topological crystalline insulators (TCI) \cite{Fu11, Hsieh12} in IV-VI semiconductors SnTe \cite{Tanaka12} and Pb$_{1-x}$Sn$_x$Se(Te) \cite{Dziawa12, Xu12a, Tanaka13} provide a new platform for topological band gap engineering via elastic strain. In this work, by first-principles calculations we demonstrate a new approach to effective band gap engineering by controlling elastic strain and membrane thickness in TCI membranes, which is based on the unique electronic properties of TCI surface states. Specifically, we find that strain induces a wave vector (${\bf k}$) shift of surface-state Dirac cones and thereby continuously changes the penetration depth of surface states. In a free-standing TCI membrane with two surfaces, the above will cause an exponential change in the hopping integral between the top and bottom surface states and thus drastically modulate the hybridization gap from 0 to 0.5 eV. This thickness and elastic strain dependent two-dimensional band gap leads to potential applications in piezoresistive devices, tunable infrared detectors, and thermoelectrics. In particular, we propose a frequency-resolved infrared photodetector using force-concentration induced inhomogeneous elastic strain in TCI nanomembranes. 

While our main results are applicable to TCI in IV-VI semiconductor family, for concreteness we choose SnTe as a representative material. As shown later, the electronic structure change in SnTe nanomembrane under elastic strain is a direct consequence of the corresponding change in infinite and the semi-infinite SnTe, therefore our calculations and analyses will address three progressively more complex geometries: (a) infinite bulk with 3D band structure, (b) semi-infinite bulk with a single free surface and 2D surface band structure, and (c) nanomembrane with two coupled surfaces and 2D band structure, respectively. 

\section{Strain-dependent electronic structure of bulk, semi-infinite, and nanomembrane S\lowercase{n}T\lowercase{e} }

\subsection{Electronic structure in bulk S\lowercase{n}T\lowercase{e}}
\label{bulk}

\begin{figure}[htbp]
\centering
\includegraphics[width=1\columnwidth]{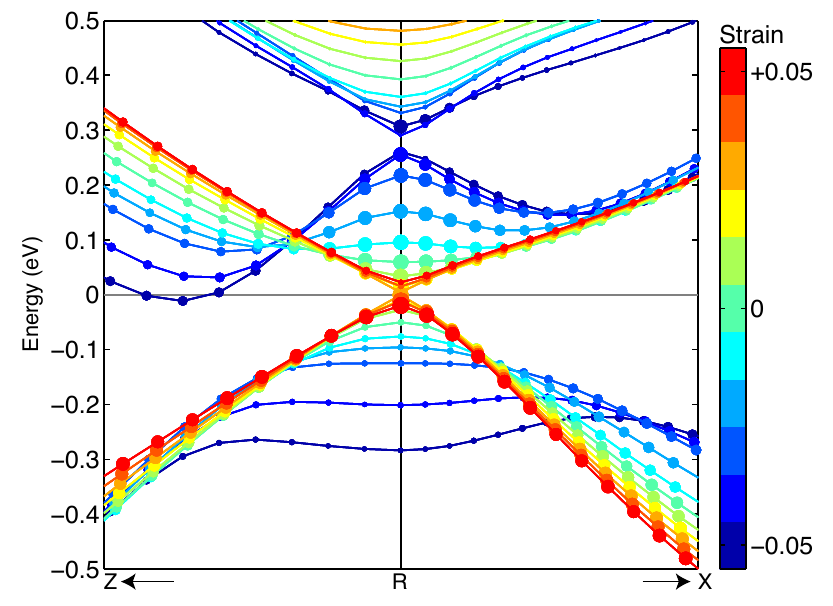}
\caption{Strain-dependent electronic band structure of bulk SnTe. The dot size indicates the strength of Te's contribution to the density of states.}
\label{fig1}
\end{figure}

The fundamental band gap of bulk SnTe is located at the L points of 3D Brillouin zone (BZ). Most importantly, there exists band inversion at these L points at strain-free condition, i.e. the anions dominate the lowest conduction band while the cations dominate the highest valence band, which is a necessary condition for realizing the TCI phase and creating topological surface states \cite{Hsieh12}. Using first-principles density-functional theory (DFT) calculations \cite{Hohenberg64, Kohn65}, we calculated bulk band structure of SnTe near L as a function of in-plane ($x$-$y$) biaxial strain with atomic structure fully relaxed along $\hat{z}$. The calculated 3D band structure is shown in Fig.~\ref{fig1} whose  first BZ (purple tetragonal cell) and atomic configuration are shown in Fig.~\ref{fig-S1}(a) and \ref{fig-S1}(b) of the Supplemental Material, respectively (see the Supplemental Material). The size of the dots in Fig.~\ref{fig1} indicates the contribution of Te's $s$ and $p$ orbitals to the corresponding electronic state. As the strain increases, the band gap decreases, and then opens up again. Above 3\% strain, band inversion disappears, that is, Te's $p$ orbitals have shifted from the conduction band edge to the valence band edge. Such strain-tunable band inversion and band gap in {\it bulk} SnTe provide the basis of strain-tunable band gap of two-dimensional SnTe membrane to be described below.  

\subsection{Metallic surface states in semi-infinite S\lowercase{n}T\lowercase{e}}
\label{semiinfinite}
In order to address the strain- and thickness-dependent band gap in TCI membranes, it is necessary to first characterize the metallic TCI surface state in the semi-infinite limit, where we study the low-energy electronic structure of a single isolated surface. Tight-binding Hamiltonian of semi-infinite SnTe surface is constructed in the first-principles quasiatomic Wannier function basis \cite{Qian08, Lu04, Marzari12} by knitting bulk SnTe Hamiltonian and surface Hamiltonian. The latter was cut from a fully relaxed 24-layer SnTe slab. A self-energy correction was further applied to bulk Hamiltonian by half-surface Green's function using a highly convergent algorithm \cite{Sancho84, Sancho85} often adopted in quantum transport calculations \cite{Qian10}. The calculated band structure along $\overline{\Gamma}$$\overline{\mathrm X}$ for a top surface (+$\hat{z}$) is shown in Fig.~\ref{fig2}(a) whose surface BZ is shown in Fig.~\ref{fig-S1}(a) of the Supplemental Material (blue square). Two chiral Dirac cones with the same chirality emerge at ($E$, $k_x$, $k_y$) of ($E_F$, 0.66 \AA$^{-1}$, 0) and ($E_F$, 0.73 \AA$^{-1}$, 0), respectively. In the vicinity of the cone almost all the spin moment comes from $\langle {\bf S} \rangle_y$ with its direction marked in Fig.~\ref{fig2}(a): red for positive $\langle {\bf S} \rangle_y$ and blue for negative  $\langle {\bf S} \rangle_y$. The estimated Fermi velocity at the left Dirac cone is 2.8$\times$10$^5$ m/s for red band and 1.6$\times$10$^5$ m/s for blue band. The asymmetric Fermi velocity results from different dominating components in two bands, that is, Sn-dominated red band and Te-dominated blue band at the left Dirac cone. In addition, energy contour and spin texture of surface states at conduction band minimum (CBM), CBM+1, valence band maximum (VBM), and VBM-1 in 2D BZ are presented in Fig.~\ref{fig2}(b-e), revealing a clear anisotropy of both the shape of Dirac cones and the strength of spin moment  $\langle {\bf S} \rangle$ in $k_x$ and $k_y$ direction.

\begin{figure*}[htbp]
\centering
\includegraphics[width=0.9\textwidth]{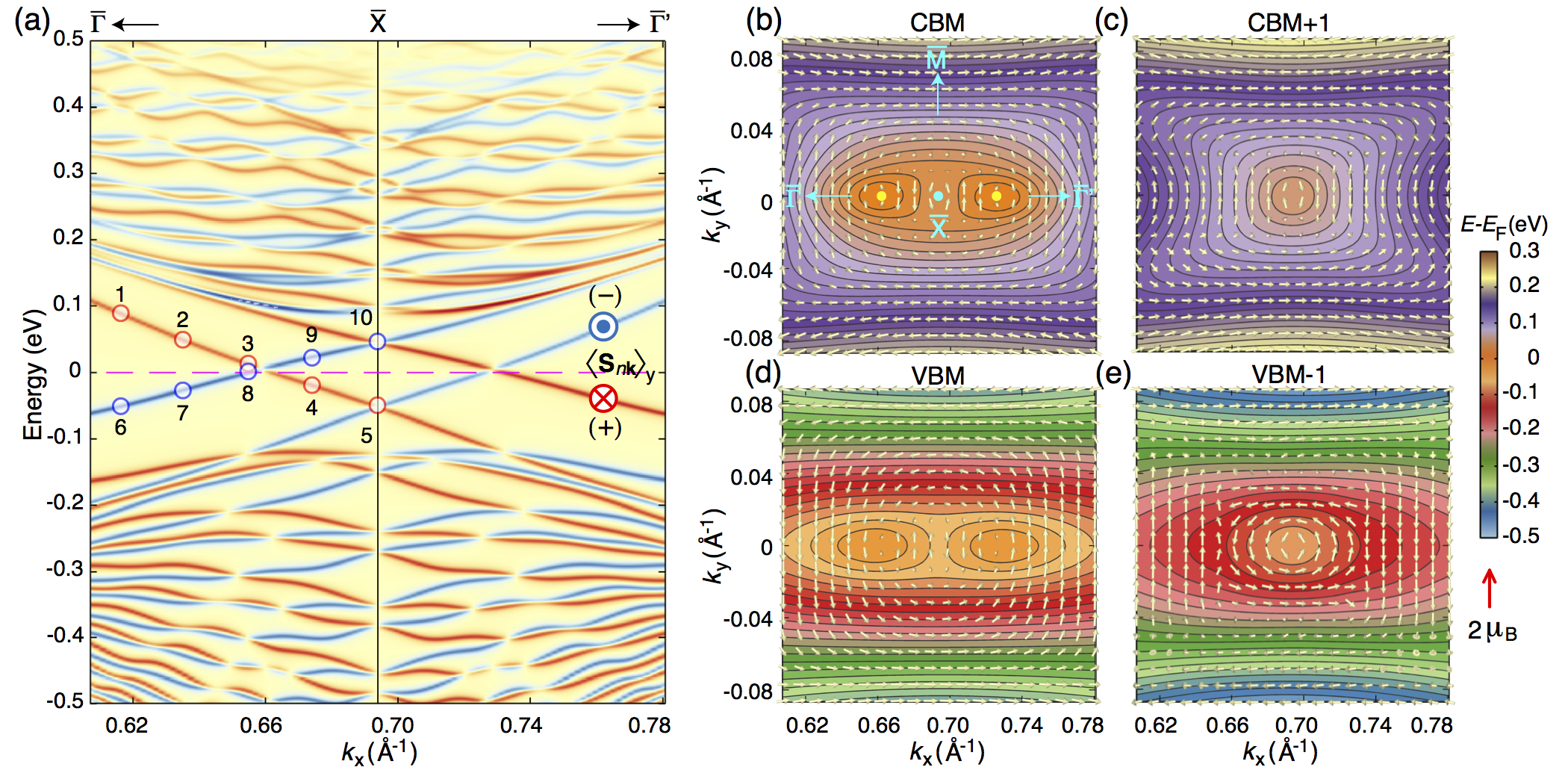}
\caption{Band structure and spin texture of strain-free semi-infinite SnTe. (a) Band structure of strain-free semi-infinite SnTe color-encoded by spin component along $\hat{y}$. Vacuum is in the +$\hat{z}$ direction. (b-e) Energy contour and spin texture of surface states at CBM, CBM+1, VBM, and VBM-1 in 2D Brillouin zone.}
\label{fig2}
\end{figure*}

\subsection{Strain and membrane thickness dependent band gap in S\lowercase{n}T\lowercase{e} nanomembranes}
\label{nanomembrane}

To further reveal the spatial profile and atomic orbital composition of surface states, we present in Fig.~\ref{fig-S2} (see the Supplemental Material) the atom-projected local density of states (LDOS) of ten selected surface states near the left Dirac cone with their positions marked in Fig.~\ref{fig2}(a). The estimated decay length increases as the 2D BZ ${\bf k}$-point moves towards $\overline{\mathrm X}$. This is because $\overline{\mathrm X}$-point of the 2D BZ contains a projection of R(L)-point in 3D BZ; as the surface state ${\bf k}$-point in 2D BZ approaches $\overline{\mathrm X}$, it comes into energy resonance with bulk band state and is then no longer localized in $\hat{z}$. The penetration depth and the amplitude of the atom-projected LDOS also highly depend on specific type and site of atoms. More details are given in the Supplemental Material. 

\begin{figure*}[htbp]
\centering
\includegraphics[width=0.8\textwidth]{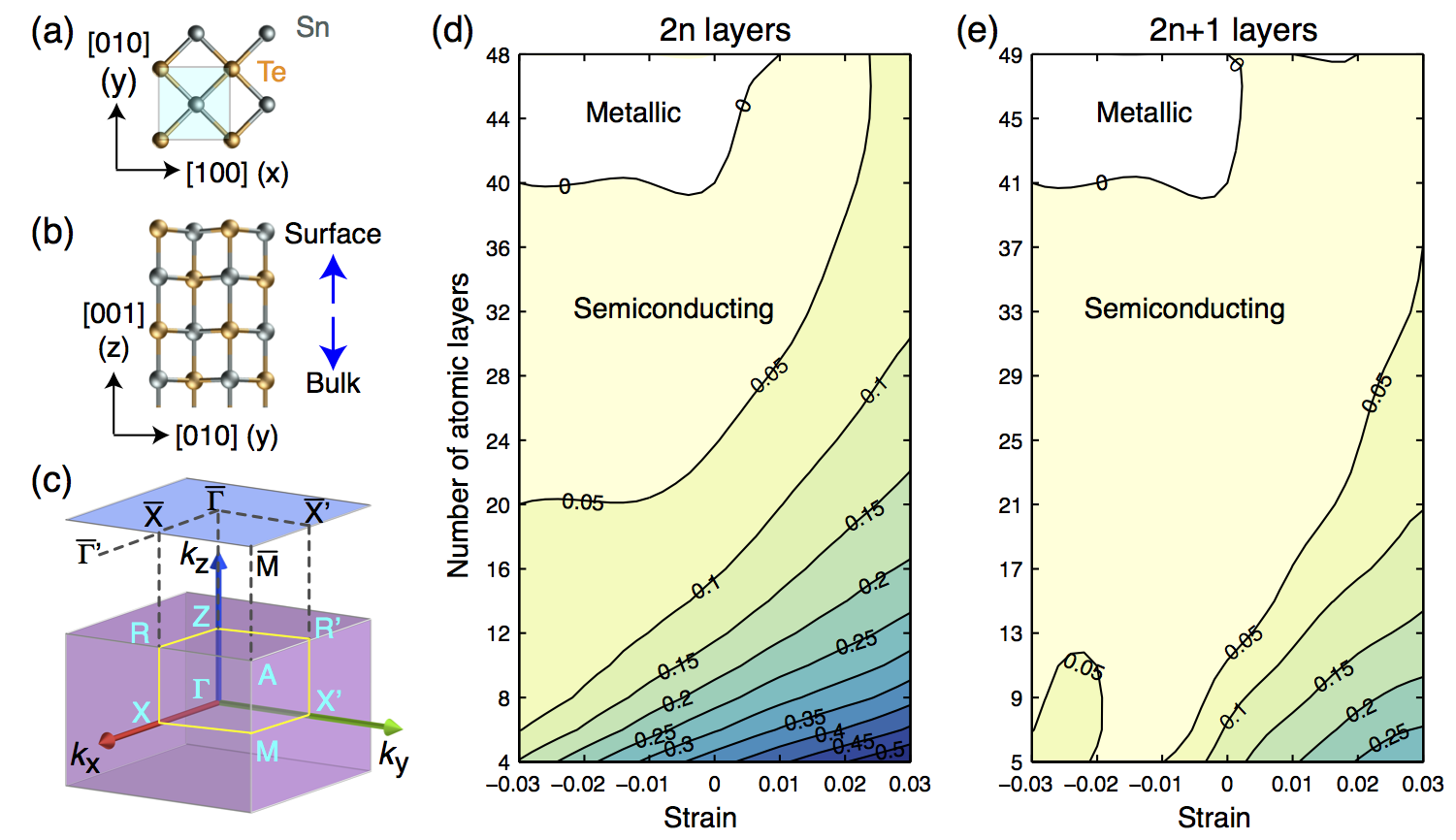}
\caption{Strain and thickness dependent band gap in SnTe nanomembrane. (a) Top and (b) side view of equilibrium atomic structure of SnTe membrane. Shaded area in cyan represents the unit cell for first-principles calculations. (c) Projection of 3D Brillouin zone (purple) for bulk SnTe onto 2D surface Brillouin zone (blue) for SnTe membrane. (d) Biaxial-strain dependent band gap in membranes with even number of atomic layers. (e) Biaxial-strain dependent band gap in membranes with odd number of atomic layers.}
\label{fig3}
\end{figure*}

Due to the finite penetration depth of the above isolated surface states, when the thickness of SnTe membrane decreases to tens of layers, the top and bottom surface states will hybridize with each other, which creates a finite band gap \cite{Liu13b}. To demonstrate this hybridization gap and reveal its strain dependence, we built fourteen different slab configurations of SnTe with the number of atomic layers varying from 4 to 49. The slabs sit on the $x$-$y$ plane as shown in Fig.~\ref{fig3}(a) with its unit cell indicated as the blue area containing four atoms. In-plane biaxial strain from -3\% to +3\% was subsequently applied to each strain-free SnTe membrane of varying thickness, which preserves the mirror symmetry of SnTe (001) surface with respect to the (110) plane. Geometry optimization leads to the small relaxation of Sn and Te atoms along $\hat{z}$ shown in Fig.~\ref{fig3}(b). We find that Sn atoms near the surface pop out of the $x$-$y$ plane towards the vacuum while Te atoms contract towards the bulk of the membrane.  As illustrated in Fig.~\ref{fig3}(c), the first surface BZ (blue square)  is simply the projection of the first bulk BZ (purple tetragonal cell) onto the (001) plane. As a result, the high symmetric points ($\Gamma$, Z), (X, L), (X$'$, L$'$) and (M, A) in the 3D BZ are projected to $\overline{\Gamma}$, $\overline{\mathrm X}$, $\overline{\mathrm X'}$, and $\overline{\mathrm M}$ in the surface 2D BZ. 

Figure~\ref{fig3}(d) and \ref{fig3}(e) show the strain- and thickness-dependent band structure of SnTe membranes. At a fixed number of SnTe atomic layers, the direct band gap $E_g (\Lambda)$  of nanomembrane increases as the strain increases (note this is {\it opposite} to the trend in bulk SnTe, for small tensile strains); while at a fixed strain, the band gap $E_g (\Lambda)$  increases as the thickness reduces due to stronger hybridization effect in thinner membranes. In particular, SnTe nanomembrane with about 40 atomic layers ($\sim$12.5 nm) can be tuned from zero gap to a finite gap when biaxial strain sweeps from -3\% to 3\%. The tunability of the nanomembrane band gap via elastic strain is significant. For example, the band gap of 16-layer SnTe nanomembrane increases by more than 50\% from 0.1 eV to 0.15 eV under 1\% strain. With increasing biaxial strain and decreasing membrane thickness, the band gap varies from 0 to 0.5 eV. The largely tunable bandgap from metal to insulator has potential applications in thin-film piezoresistive devices, infrared sensors and thermoelectrics. 

Furthermore, we find a striking difference in the electronic structures of TCI membranes with an odd and even number of atomic layers, shown in Fig.~\ref{fig3}(d) and \ref{fig3}(e) respectively. A better illustration of this even-odd oscillation is shown in Fig.~\ref{fig-S3}(a) of the Supplemental Material. This unexpected behavior originates from the different crystal symmetry in the slabs with even and odd number of layers. Membranes with an odd number of layers possess the reflection symmetry $z \rightarrow -z$ with respect to the atomic layer in the middle \cite{Liu13b}, while membranes with an even number of layers possess a two-fold screw axis. The consequence is that TCI membranes with an even number of layers result in larger gap at the zone boundary $\overline{\mathrm X}$ and $\overline{\mathrm X'}$, thus larger gap at the Dirac cones. More details are given in the Supplemental Material. 

\section{Origin of strain-induced electronic phase transition in TCI S\lowercase{n}T\lowercase{e} nanomembranes}
\label{origin}
\begin{figure*}[htbp]
\centering
\includegraphics[width=1\textwidth]{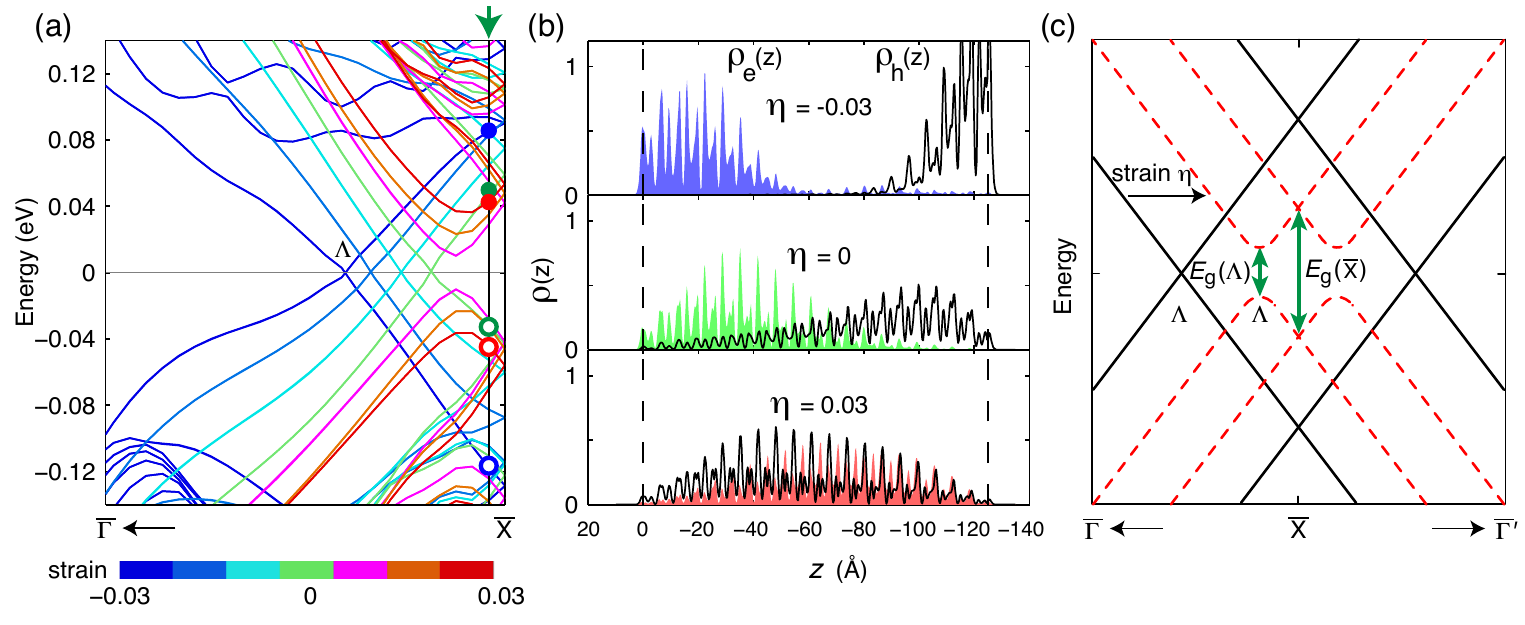}
\caption{Strain-induced electronic phase transition for 40-layer SnTe nanomembrane. (a) First-principles electronic band structures of 40-layer SnTe nanomembrane under biaxial strain $\eta$ from -3\% to +3\%. (b) Strain-dependent charge density profiles of CBM (shaded area) and VBM (black curves) surface states close to $\overline{X}$, corresponding to the electrons and holes marked as filled and unfilled dots in (a). The charge density profiles are obtained from first-principles calculations and integrated along $x$-$y$ plane parallel to the membrane surface. (c) Schematic of strain-induced electronic band structure change of SnTe nanomembrane.}
\label{fig4}
\end{figure*}

To understand the origin of the above strain-induced band gap change and gap closing/opening, we calculated the first-principles strain-dependent band structure of a 40-layer SnTe membrane displayed in Fig.~\ref{fig4}(a). It clearly shows the finite band gap $E_g (\Lambda)$  (a quantity for SnTe nanomembrane) under the biaxial strain of above 1\%. 

The VBM and CBM states in nanomembrane close to $\overline{\mathrm X}$ are presented in Fig.~\ref{fig4}(b). A schematic is shown in Fig.~\ref{fig4}(c) where the gap is indicated by $\Lambda$ which is close to the zone boundary $\overline{\mathrm X}$. As biaxial strain increases, the Dirac cone $\Lambda$ shifts towards $\overline{\mathrm X}$ and the band gap $E_g (\Lambda)$ increases. Its origin can be understood as follows. Biaxial tensile strain causes an increase in the lattice constant, which reduces the band inversion strength in bulk SnTe at R (L) \cite{Hsieh12}. Because R (L) in bulk BZ is projected onto $\overline{\mathrm X}$ in the surface BZ (Fig.~\ref{fig-S1} of  the Supplemental Material), the band inversion strength in semi-infinite SnTe at $\overline{\mathrm X}$ also linearly decreases, resulting in the decreasing energy gap between VBM and CBM at $\overline{\mathrm X}$ in semi-infinite SnTe, labeled as $E_g^S (\overline{\mathrm X})$. The $E_g (\overline{\mathrm X})$ of SnTe nanomembrane approaches $E_g^S (\overline{\mathrm X})$ when the membrane is thick enough, and both have the same trend with strain, as does $E_g(\Lambda)$: all three quantities {\it decreases with tensile strain}. This can be verified by comparing Fig.~\ref{fig2}(a) and Fig.~\ref{fig4}(a). Since the position of Dirac cone $\Lambda$ is determined by the intersecting point of inverted bands \cite{Liu13b, Liu13a}, the cone of the nanomembrane then moves closer to $\overline{\mathrm X}$ as $E_g^S (\overline{\mathrm X})$ and $E_g (\overline{\mathrm X})$ decrease under increasing strain. As demonstrated in Fig.~\ref{fig-S2} of the Supplemental Material, the penetration length of the surface states strongly depends on the 2D BZ ${\bf k}$-point, hence the closer the Dirac point $\Lambda$ to $\overline{\mathrm X}$, the longer the penetration length. Therefore, under increasing strain, the $E_g(\Lambda)$ / $E_g^S (\overline{\mathrm X})$-reduction-induced cone shift leads to an increase in the penetration length of surface state in semi-infinite SnTe. This gives rise to an {\it exponential} increase in the hybridization strengths of two TCI surface states at opposite sides of SnTe nanomembrane, and thus dramatically {\it increases} the hybridization gap $E_g(\Lambda)$ of the nanomembrane. In contrast, compressive strain reduces the surface state penetration length \cite{Barone13}. For thinner SnTe membrane, the finite intrinsic penetration depth $\lambda_0$ even at zero strain causes a large hybridization gap even with no strain, and it remains finite under small strain from -3\% to 3\%. From Fig.~\ref{fig4}(a), the estimated average ${\bf k}$-shift of the Dirac point $\Lambda$ along the [100] direction in the 40-layer SnTe is +0.0143 \AA$^{-1}$ per 1\% biaxial-strain, while the corresponding change of $E_g (\overline{\mathrm X})$ is about 16 meV per 1\% biaxial-strain. The strain tunable band gap in TCI thin film and its underlying mechanism described above are the main results of our work. The significance and implication of the strain-dependent Dirac cone shift are discussed later.

\section{Electronic hybridization induced semiconducting surface states in S\lowercase{n}T\lowercase{e} membrane}
\label{hybridization}
\begin{figure*}[htbp]
\centering
\includegraphics[width=0.8\textwidth]{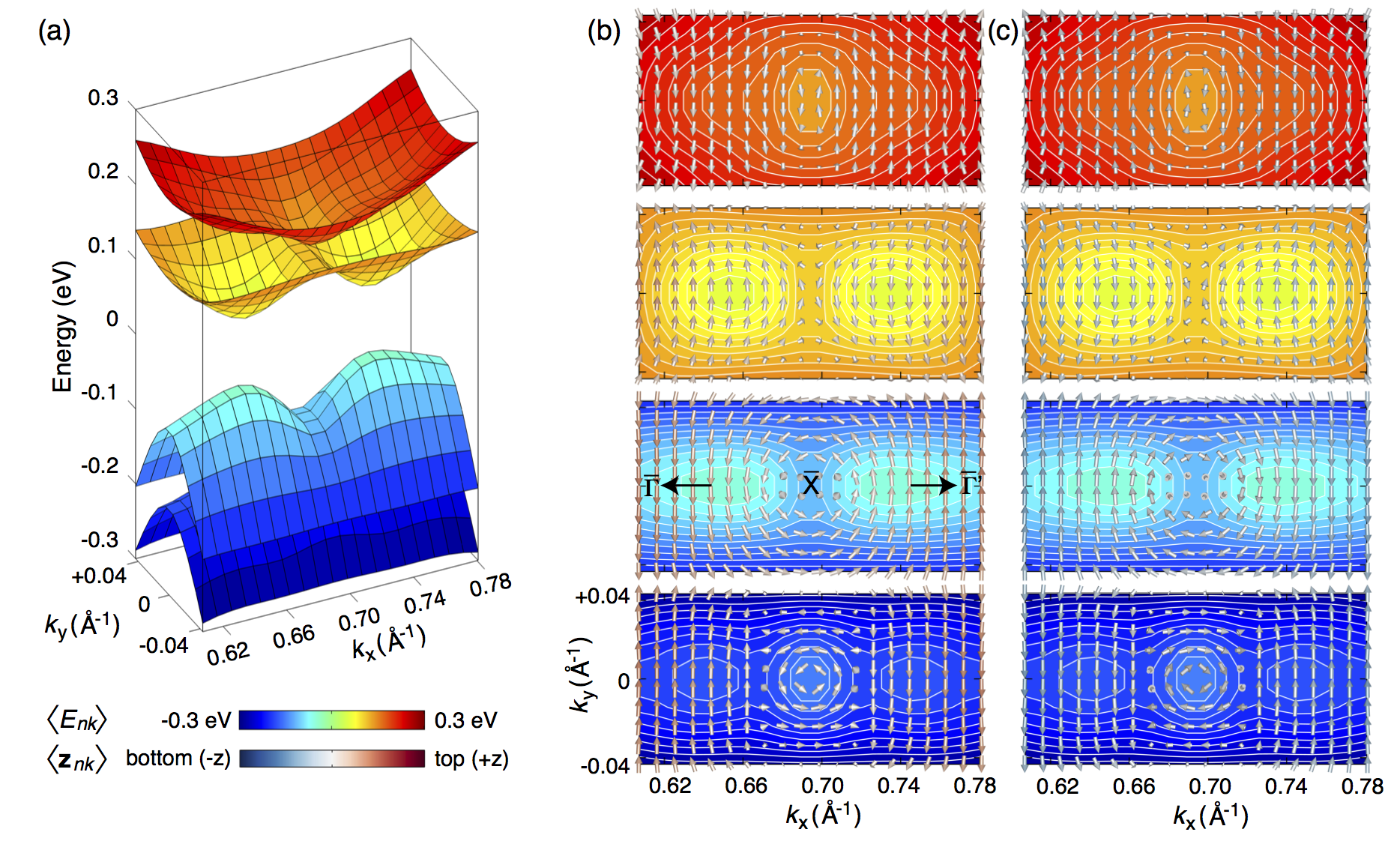}
\caption{Band structure of a 16-layer SnTe nanomembrane at strain-free condition. (a) Energy dispersion of eight electronic bands close to the Fermi level in 2D Brillouin zone around $\overline{\mathrm X}$. Due to double-degeneracy only four surfaces are distinguishable in (a). These eight bands are separated into two groups in (b) and (c) according to the charge center position $\langle z_{n {\bf k}} \rangle$ in the $\hat{z}$ direction. (b) Energy contour and spin texture of the first group with their charge center position close to the top surface. (c) Energy contour and spin texture of the second group with their charge center position close to the bottom surface. The arrows in (b) and (c) for spin textures are colored by the charge center position relative to the top and bottom surface.}
\label{fig5}
\end{figure*}

The symmetry-protected spin-momentum locking in the surface states can help reduce back scattering in carrier transport, enhance carrier mobility, and thus lower joule heating. It is particularly useful for designing electronic and optoelectronic devices with reduced dissipation. A natural question is: how does the electronic structure change prior to the emergence of massless Dirac cone with increasing membrane thickness? We plot in Fig.~\ref{fig5}(a) the energy dispersion of a 16-layer strain-free SnTe membrane which is thin enough to have a finite band gap $E_g(\Lambda)$, yet thick enough to allow us to distinguish top and bottom surface states. Former massless Dirac cones near the Fermi level are now separated by a finite energy gap, forming two massive Dirac cones. Due to double degeneracy, only four surfaces near the Fermi level are distinguishable from the energies in Fig.~\ref{fig5}(a). Hence to reveal the spin texture near the massive cones, we separate the double-degenerate surfaces into two groups according to the charge center in the $\hat{z}$ direction $\langle z_{n {\bf k}} \rangle$. Band structure and spin texture of the bands with their charge center close to the top (bottom) surface are displayed in Fig.~\ref{fig5}(b) (Fig.~\ref{fig5}(c)) with spin texture indicated by arrows and color-coded by $\langle z_{n {\bf k}} \rangle$. The original chiral texture close to the massless Dirac cones in Fig.~\ref{fig5} cannot be clearly distinguished in Fig.~\ref{fig5}(b) and 5(c) as $\langle z_{n {\bf k}} \rangle$ of the corresponding doubly degenerate eigenstates are very close to each other (near the center of the 16-layer slab). Nonetheless, away from the massive Dirac cones (e.g., towards $\overline{\Gamma}$ and $\overline{\Gamma'}$) the large in-plane spin moments exhibit a similar pattern as Fig.~\ref{fig4}, and their charge center $\langle z_{n {\bf k}} \rangle$ becomes closer and closer to the top and bottom surfaces, which is consistent with the trend in the spatial penetration of surface states at massless Dirac cone displayed in Fig.~\ref{fig-S2} of the Supplemental Material. 

Interestingly, Figure~\ref{fig5}(b) and \ref{fig5}(c) demonstrate the existence of in-plane chiral spin texture at $\overline{\mathrm X}$ that is similar to the case in Fig.~\ref{fig2}. This robust chiral texture at $\overline{\mathrm X}$ against the membrane thickness reduction is due to the two-fold screw rotation mentioned above. The similar in-plane spin textures around $\overline{\mathrm X}$ in both massless and massive Dirac cones indicate that the spin-momentum texture happens before the emergence of massless Dirac cone, thus the back-scattering of carriers in semiconducting SnTe membranes should also be reduced significantly compared to ordinary semiconductors with no spin texture. 

\section{Applications}
\label{applications}
\begin{figure}[htbp]
\centering
\includegraphics[width=1\columnwidth]{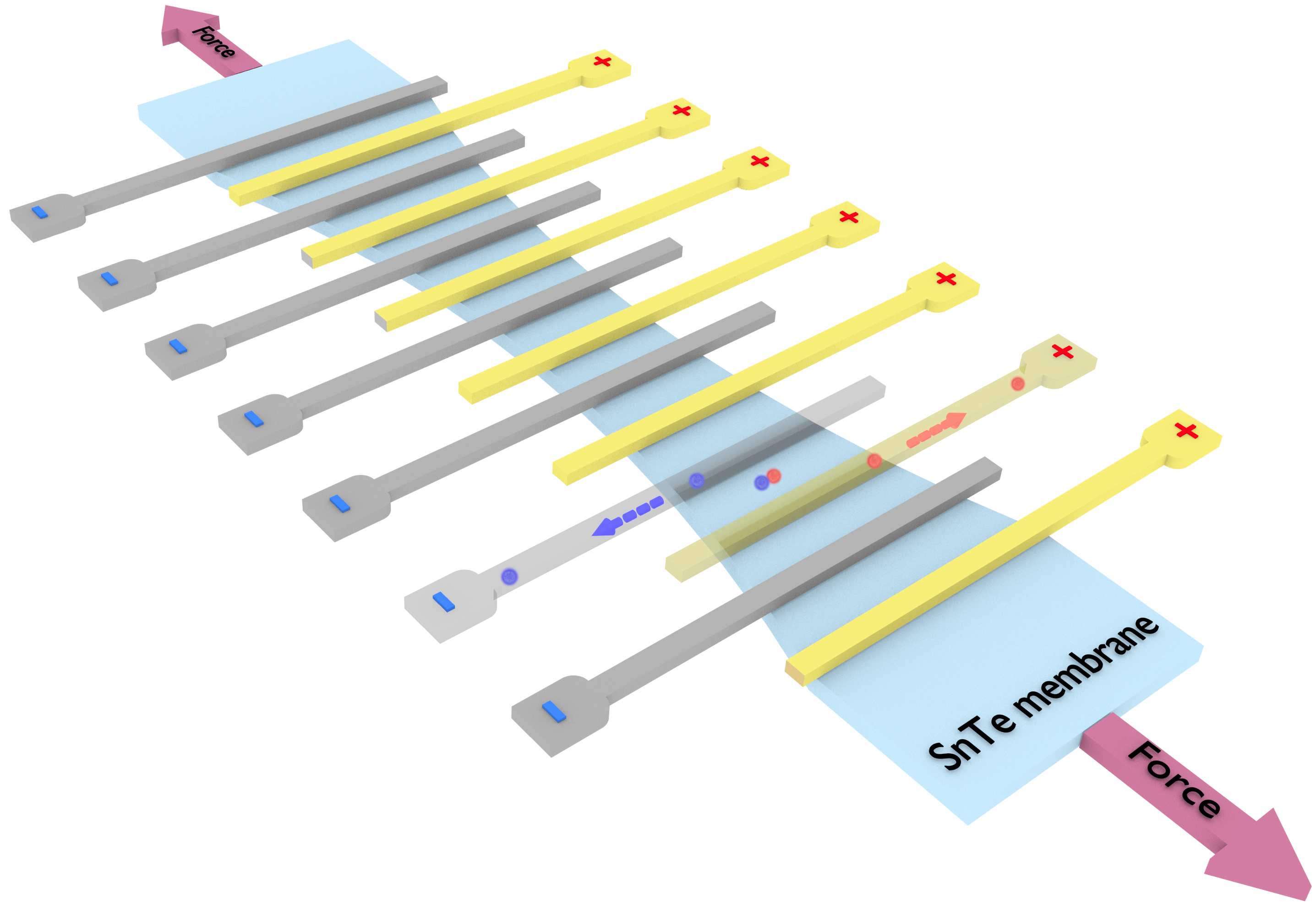}
\caption{Frequency-resolved infrared photodetector using inhomogeneously strained SnTe membrane. The spatially varying of SnTe membrane induces the inhomogeneous strain. The narrower membrane ribbon, the higher elastic strain, thus the higher band gap. Consequently, a spatial dependent optical band gap is achieved. The photoexcited electrons and holes will migrate towards cathode and anode, forming a measurable photocurrent from the local region. Depending on specific photon energy of interest, one may choose a suitable material, thickness, geometry, and force.}
\label{fig6}
\end{figure}

The strain tunable band gap of SnTe membrane enables us to design frequency-resolved ultrathin infrared photodetector. A schematic of the setup is shown in Fig.~\ref{fig6}. A SnTe nanomembrane with spatially varying width is in contact with two types of electrodes exhibiting different work functions. Because of the inhomogeneous strain induced by force concentration \cite{Feng12}, the optical band gap is spatially varying: the narrower the membrane ribbon, the higher elastic strain, thus the higher band gap. This allows for the detection of multi-energy photons within a single device. The photoexcited electrons and holes will migrate towards cathode and anode, forming a measurable photocurrent from the local region. Depending on specific photon energy of interest, one may choose a suitable material, thickness, geometry, and force. 

\section{Conclusions}
\label{conclusions}
The mechanism of strain tunable band gap proposed in this work is unique to TCI. For comparison, in $Z_2$ topological insulators, while the penetration depth of surface states in semi-infinite $Z_2$ TIs can be tuned by strain, the Dirac cone is pinned at time-reversal-invariant momenta, contrary to the strain-induced shift in TCI. On the other hand, while Dirac cone ${\bf k}$-shift is known to occur in graphene \cite{Pereira09a, Pereira09b, Guinea10, Levy10}, it does not create a band gap as the interior is always gapless. Last but not the least, electronic states in TCI nanomembranes exhibit spin textures, which may be useful for reducing dissipation in applications such as infrared detection and piezoresistive sensing. In addition, elastic strain provides a way to engineer both topologically protected metallic and strong spin-textured semiconducting surface states within a single material. For example, one can elastically stretch the middle part of SnTe membrane ribbon while leaving the rest stress free, resulting in a metal-semiconductor-metal pseudo-hetero structure within a single SnTe membrane. Such inhomogeneous strain-modulated pseudoheterostructure \cite{Feng12, Nam13} may have applications in novel electronic and optoelectronic devices.

SnTe nanomembranes belong to the class of ultrastrength materials \cite{Zhu10, Zhu09}, thus are expected to be able to sustain a large elastic strain before plasticity or fracture takes place. As free-standing ultrathin silicon nanomembranes have been synthesized and demonstrated \cite{Roberts06}, we expect that ultrathin SnTe membranes can also be synthesized and the elastic strain and membrane thickness controlled metal-to-semiconductor transition in SnTe nanomembranes can be experimentally verified, by for example optical interrogations.

In conclusion, we have demonstrated the possibility to tailor electronic band gap and electronic phase transition of SnTe membrane by varying its thickness and applying elastic strain, and achieved a tunable band gap varying from 0 to 0.5 eV. Our results also reveal a striking even-odd oscillation behavior of band gap, the physical origin of which roots in two-fold screw rotation symmetry preserved in membranes with even number of layers, but absent in the case of odd number of layers. We also found strong spin-textured semiconducting surface states associated with massive Dirac cone for thin membranes, complementary to topologically protected metallic surface states with massless Dirac cone for thicker membranes. Membrane thickness control and elastic strain engineering of band gap are generally applicable to other topological insulators. However, elastic strain cannot introduce Dirac cone shift in $Z_2$ TI membrane as the Dirac cone is pinned at time-reversal-invariant momenta. In contrast, strain can introduce a 2D cone shift in TCI surface state even with the crystalline symmetry protection, making the TCI nanomembrane a unique platform to engineer a pseudo-magnetic field which may directly couple to the spin-texture of the surface states in both metallic and semiconducting state. The present findings will enable the applications of topological (crystalline) insulator nanomembrane in piezoresistive devices, thermoelectrics, and infrared detectors as well as energy-efficient electronic and optoelectronic devices. 

\begin{acknowledgments}
 This work was supported by NSF DMR-1120901 (X.Q. and J.L.) and U.S. Department of Energy, Office of Basic Energy Sciences, Division of Materials Sciences and Engineering under Award DE-SC0010526 (L.F.). Computational time on the Extreme Science and Engineering Discovery Environment (XSEDE) under the grant number TG-DMR130038 and TG-DMR140003 is gratefully acknowledged.
\end{acknowledgments}

\appendix
\section{Methods}
\label{methods}

\subsection{Atomic and electronic structure of SnTe bulk and membrane}

First-principles DFT calculations were performed using the Vienna {\it ab initio} simulation package (VASP) \cite{Kresse96b, Kresse96a}. We employed projector augmented-wave (PAW) method \cite{Bloechl94}, the generalized-gradient approximation (GGA) \cite{Langreth83, Becke88} of exchange-correlation functionals in the Perdew-Berke-Ernzerhof (PBE)'s form \cite{Perdew96}, energy cutoffs of 220 eV for plane-wave basis and 380 eV for augmented charge density, and 6$\times$6$\times$1 Monkhorst-Pack ${\bf k}$-point sampling \cite{Monkhorst76}. Fourteen slab configurations of SnTe sitting on the $x$-$y$ plane were built using the relaxed bulk lattice constant with number of atomic layers varying from 4 to 49. Each configuration was further optimized with maximum energy fluctuation less than 10$^{-6}$ eV, and maximum residual force less than 0.02 eV/\AA. In-plane biaxial strain from -3\% to +3\% was subsequently applied to each strain-free SnTe slab of different thickness. To minimize the periodic image interaction from periodic boundary condition in the calculations, a large vacuum region of more than 20 {\AA}  was added along the surface normal. The subsequent electronic band structure for different biaxial strain is obtained from DFT-PBE calculations. 

\subsection{Electronic structure of semi-infinite S\lowercase{n}T\lowercase{e}}

Tight-binding Hamiltonian of semi-infinite SnTe surface is constructed in the first-principles quasiatomic Wannier function basis \cite{Qian08, Lu04, Marzari12} by knitting bulk SnTe Hamiltonian and surface Hamiltonian cut from a fully relaxed 24-layer SnTe slab. An energy shift is introduced to the on-site energy of bulk Hamiltonian to maintain charge neutrality. The Hamiltonian of semi-infinite SnTe surface was formed by a slab of finite thickness with its Hamiltonian component in the coupling region adjusted by a frequency-dependent self-energy correction. The self-energy correction is calculated from half-surface Green's function using a highly convergent algorithm \cite{Sancho84, Sancho85} often adopted in quantum transport calculations \cite{Qian10}. As our main interest is near the mirror-symmetry protected Dirac cone (e.g. within the range of -0.2 to 0.2 eV), we approximate the frequency-dependent self-energy by a constant self-energy matrix calculated at Fermi level. This approach allows us to directly diagonalize the self-energy-corrected slab Hamiltonian and extract the surface eigenstates.

\section{Supplemental Materials}
\label{supplemental}

\subsection{Real and reciprocal space structures of S\lowercase{n}T\lowercase{e} bulk and membrane in different unit cells.}

\begin{figure}[h]
  \centering
  \includegraphics[width=1\columnwidth]{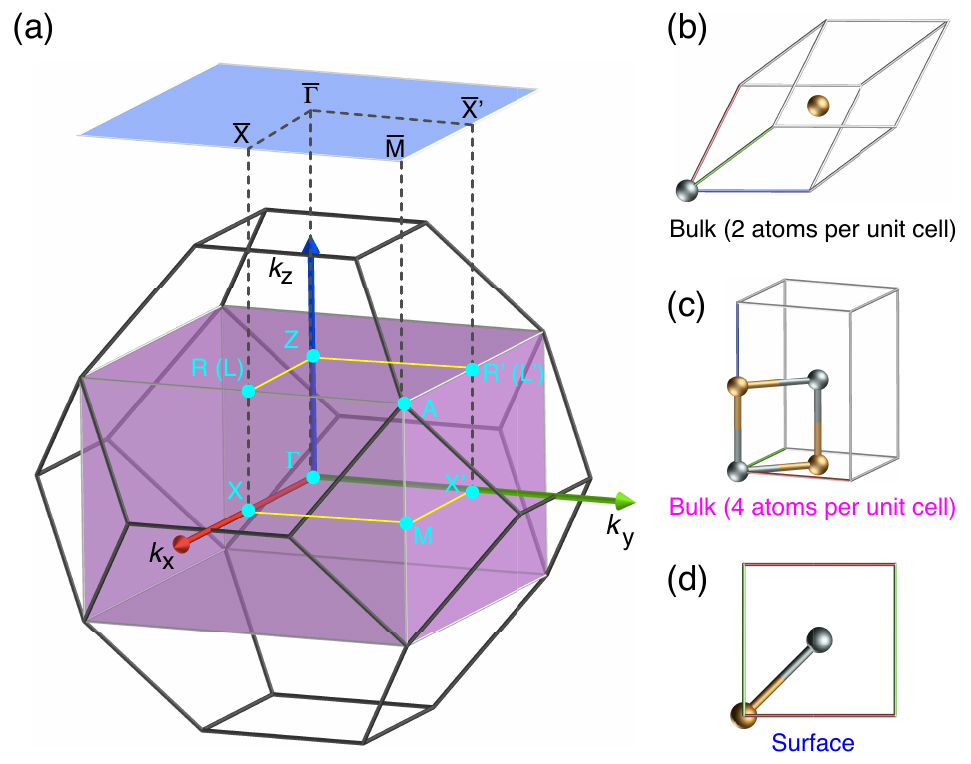}
\caption{Real and reciprocal space structures of SnTe bulk and membrane in different unit cells.  (a) Brillouin zones for bulk and SnTe membrane in different unit cells. The black one is Brillouin zone for bulk SnTe with two atoms per unit cell with its unit cell shown in (b). The purple one is for bulk SnTe with four atoms per unit cell with its unit cell shown in (c). The 2D blue one is for both SnTe semi-infinite surface and SnTe nanomembrane with its top view shown in (d).}
\label{fig-S1}
\end{figure}

\subsection{Decaying behavior of surface states}

\begin{figure}[h]
  \centering
  \includegraphics[width=1\columnwidth]{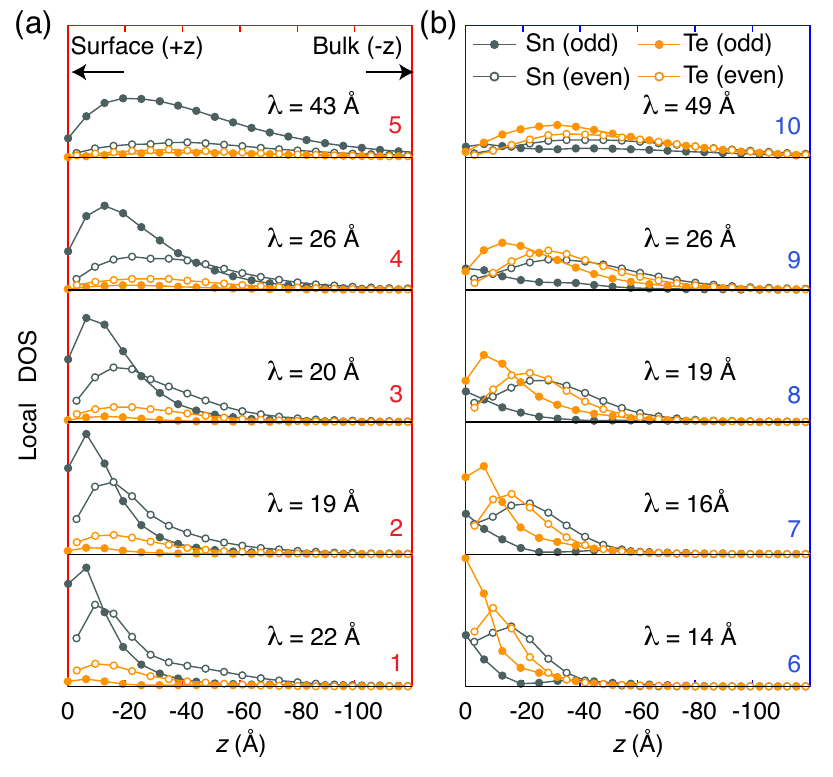}
\caption{Atom-projected local density of states of 10 selected surface states near the Dirac cone. The corresponding positions of the selected states are marked in the band structure shown in Fig.~\ref{fig2}(a). (a) for states 1-5 dominated by Sn, and (b) for states 6-10 dominated by Te. The selected k-points are located at 0.617, 0.636, 0.655, 0.675, and 0.694 \AA$^{-1}$, and the last ${\bf k}$-point is also where zone boundary $\overline{\textrm X}$ is. The decay length $\lambda_{n{\bf k}}$ is also listed for each surface state. Odd and even here indicate the position of specific atoms. The top surface layer is denoted as Layer 1.}
\label{fig-S2}
\end{figure}

To reveal the spatial profile and atomic composition of surface states, we present in Fig.~\ref{fig-S2} atom-projected local density of states (LDOS) of ten selected surface states near the left Dirac cone with their positions marked in Fig.~\ref{fig2}(a). In order to extract the decay constant, we fit the layer-averaged LDOS into an exponential form, i.e., ${\textrm{LDOS}}(z,n{\bf k})= c \cdot e^{-|z-z_0|/\lambda_{n{\bf k}}}$, where $z_0$ denotes the position of the top surface layer in the $z$ direction and $\lambda_{n{\bf k}}$ is the decay constant for state $n$ at k-point. The estimated band and ${\bf k}$-point dependent decay constants $\lambda_{n{\bf k}}$ are listed in Fig.~\ref{fig2}(a). The decay length increases as the k-point moves towards $\overline{\textrm X}$. The penetration depth and the amplitude of the atom-projected LDOS also highly depend on specific type and site of atoms. In particular, the major contribution to states 1-5 (states 6-10) comes from Sn (Te), and as approaching to $\overline{\textrm X}$ the Sn (Te) atoms in odd layers become more dominating than those in even layers. These characters are also an unambiguous manifestation of band inversion, i.e., VBM and CBM of bulk SnTe are dominated by Sn and Te, respectively.

\subsection{Even-odd oscillation of band gap in S\lowercase{n}T\lowercase{e} nanomembrane}

\begin{figure*}[h]
  \centering
  \includegraphics[width=0.8\textwidth]{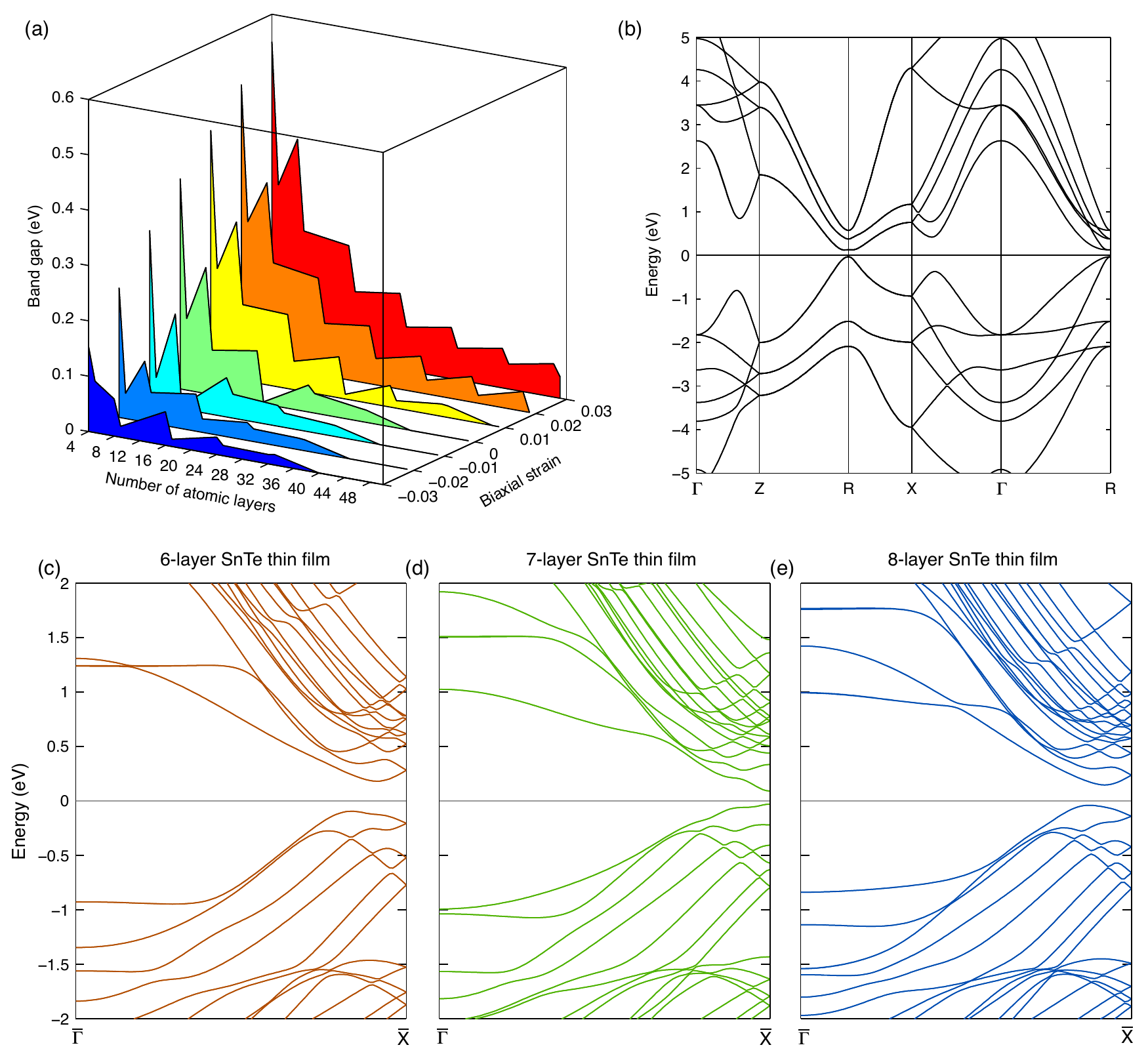}
\caption{Strain- and layer-dependent electronic structure of SnTe bulk and membrane. (a) Strain- and layer-dependent band gap of SnTe membranes with 14 different numbers of layers, i.e., 4, 5, 8, 9, 16, 17, 24, 25, 32, 33, 40, 41, 48, and 49. (b) Band structure of bulk SnTe with four atoms per unit cell at strain free condition. Its unit cell is shown in Fig.~\ref{fig-S1}(c). (c-e) Band structure of 6-layer, 7-layer, and 8-layer SnTe membrane. The degeneracy at the zone boundary $\overline{\textrm X}$ is preserved in both 6- and 8-layer case protected by two-fold screw rotation symmetry, but is broken in the 7-layer case due to the absence of this symmetry.}
\label{fig-S3}
\end{figure*}

Comparing Fig.~\ref{fig3}(d) and (e), there exists a striking behavior in the band gap which depends the odd and even number of atomic layers in the membrane. A better illustration of this even-odd oscillation is shown in Fig.~\ref{fig-S3}(a). This unanticipated behavior of band gap is in fact originated from different inherent symmetry in the slabs with even and odd number of layers. In crystalline SnTe (see the unit cell in Fig.~\ref{fig-S1}(c)), there exists two-fold screw rotation (or glide-reflection) along the [100] (also [010] and [110]) direction with the axis passing through the center of the nearest Sn-Sn bond and a displacement of 1/2 lattice vector in [100] (also [001] and [110]). Screw rotation is incompatible with Bloch theorem derived from translational invariance of crystals, thus the two-fold screw rotation naturally introduces double degeneracy for each band at the BZ boundary along [100], i.e., the $\overline{\textrm X}$ point (also Z and L). The two-fold screw-rotation-induced double degeneracy is clearly evident in the band structure of crystalline SnTe along Z-R-X in Fig.~\ref{fig-S3}(b). This screw rotation symmetry is preserved in the membrane with an even number of atomic layers, but is broken in the case of odd number of layers. The degeneracy is, therefore, preserved in the band structure of the even one, while it is lifted in the odd one. The lifted degeneracy in the membrane with an odd number of layers results in lower conduction band minimum (CBM) and higher valence band maximum (VBM) at the zone boundary $\overline{\textrm X'}$ and $\overline{\textrm X'}$. Since the band gap of the membrane in the whole 2D BZ is close to $\overline{\textrm X}$ and $\overline{\textrm X'}$, the band gap in the membrane with an odd number of layers is expected to be smaller than the one with an even number of layers. 

To give an illustration, we first establish {\it ab initio} tight-binding Hamiltonian for SnTe membrane in the bulk quasiatomic Wannier function basis transformed from first-principles Kohn-Sham eigenstates of crystalline SnTe \cite{Qian08, Marzari12,Lu04}. Its unit cell is displayed in Fig.~\ref{fig-S1}(c) and the spin-orbit interaction is included in the first-principles calculations. The calculated electronic band structures of 6-, 7- and 8-layer SnTe slabs along $\overline{\Gamma}\overline{\textrm X}$ are shown in Fig.~\ref{fig-S3}(c-e). At $\overline{\textrm X}$, energy bands of both 6- and 8-layer SnTe slabs exhibit a four-fold degeneracy with two-fold from spin degeneracy and the other two-fold from the above screw rotation symmetry. However, in the case of 7-layer SnTe slabs the former symmetry is broken while the spin degeneracy is still conserved, therefore the four energy bands are split into two sets of two-fold degenerate ones. The lifted degeneracy in the 7-layer SnTe slabs also shifts the VBM and CBM to $\overline{\textrm X}$, resulting in a direct band gap in contrast to the indirect band gap in the vicinity of $\overline{\textrm X}$ in the even layer case.

%

\end{document}